\documentclass{article}
\usepackage{spconf,amsmath,graphicx}
\usepackage{multirow}
\usepackage{booktabs}

\title{DEEP TRIPHONE EMBEDDING IMPROVES PHONEME RECOGNITION}
\name{ Mohit Yadav $^{1 \dagger}$ and  Vivek Tyagi $ ^{2 \dagger }$  \thanks{$^{\dagger}$ Equal contribution.}}
\address{ $^{1}$TCS Research New-Delhi, India  \:\: $^{2}$American Express, India}
\begin{document}
\ninept
\maketitle

\begin{abstract}

In this paper, we present a novel \textit{Deep Triphone Embedding (DTE)} representation derived from Deep Neural Network (DNN) to encapsulate the discriminative information present in the adjoining speech frames. DTEs are generated using a four hidden layer DNN with $3000$ nodes in each hidden layer at the first-stage. This DNN is trained with the tied-triphone classification accuracy as an optimization criterion. Thereafter, we retain the activation vectors ($3000$) of the last hidden layer, for each speech MFCC frame, and perform dimension reduction to further obtain a $300$ dimensional representation, which we termed as \textit{DTE}. DTEs along with MFCC features are fed into a second-stage four hidden layer DNN, which is subsequently trained for the task of tied-triphone classification.  Both DNNs are trained using tri-phone labels generated from a tied-state triphone HMM-GMM system, by performing a forced-alignment between the transcriptions and MFCC feature frames. We conduct the experiments on publicly available TED-LIUM speech corpus. The results show that the proposed DTE method provides an improvement of absolute $\mathtt{ 2.11\%}$ in phoneme recognition, when compared with a competitive hybrid tied-state triphone HMM-DNN system.
\end{abstract}
\begin{keywords}
Phoneme classification, speech recognition, deep neural networks, hidden Markov models.
\end{keywords}
\section{INTRODUCTION}
\label{sec:intro}

With the advent of deep learning, various difficult machine learning tasks such as speech recognition, image classification, and requiring Natural Language Processing (NLP) have seen notable advances in their respective performances \cite{dahl2012context,hinton2012deep,bptt}. As is well known, speech is produced by modulating a small number of parameters of a dynamical system and this implies that speech features may reside in a low dimensional manifold $\hat{X}$,  which is a non-linear subspace of original high dimensional feature space $X$. DNNs with multiple hidden layers, trained with the Stochastic Gradient Descent (SGD) over large amount of labeled data, are able to learn these manifolds much better than purely generative models such as Context Dependent tied-triphone Hidden Markov Model - Gaussian Mixture Model (CD-HMM-GMM) \cite{hinton2012deep}.

 Speech recognition is not a static pattern classification problem  and entails recognition  of a time-series of speech frames in terms of linguistic symbols such as phonemes or words. It is very well known that the probability of current phoneme or triphone or word is highly dependent on previous phonemes or triphones or words; due to pronunciation and grammatical constraints of a natural language. Traditionally, this property has been used very successfully, though only at a latter last stage of speech recognition pipeline (Viterbi decoding\cite{ney1999dynamic}), in the form of probabilistic language models.  

Unfortunately, a DNN acoustic model \cite{dahl2012context,DNNKWSTyagi}, does not harness this useful dependency information since it lacks a memory state and does not remember the triphone or phoneme classification identities of the previous few frames. On the other hand, Reccurent Neural Networks (RNN) and Long-short-term-memory (LSTM) recurrent neural network \cite{graves_hybrid,ctc,phoneme_classification},
through their explicit memory cell state, are able to retain information in long-term dependencies of the input time-series data. In particular LSTM have been more successful since their memory cell consists of several gates (forget gate, input gate, output gate) which regulate the ability to add or remove information from the previous frames and the current frame \cite{vanishing_gradient_1,phoneme_classification}.   

Despite the widespread usage of recurrent neural networks, effective training and regularization of these networks is still a non-trivial task, and continues to pose numerous challenges for research \cite{difficult_rnn}. The vanishing gradient problem is among the first known issue with the training of recurrent neural network; which is mildly mitigated by the invention of LSTM networks \cite{vanishing_gradient_1, vanishing_gradient_2}. The other challenges that arise in LSTM recurrent neural networks are -- a critical dependence on weight initialization, difficulty in optimizing and regularizing, difficulties in parallel computation due to the sequential dependence on the training, and high memory requirement of the Back-Propagation Through Time (BPTT) algorithm which is used in training recurrent networks \cite{difficult_rnn,advances_rnn_optimization,bptt}. On the other hand, feed-forward DNNs, owing to their simpler architecture, are not constrained by most of these challenges and are relatively easier to train with Stochastic Gradient Descent and random initialization of the weights \cite{glorot2011deep}. Therefore, it is desirable to develop new DNN like architecture, which combines the memorization ability of the LSTM to capture useful information present in the previous and next frames for the sequence recognition tasks such as phoneme recognition.

One of the common solution is to provide a large number of adjoining speech feature frames (MFCC)  as input to train a DNN. As reported in \cite{DNNKWSTyagi}, indeed the hybrid HMM-DNN phoneme recognition accuracy improved from $79.0\%$ to $81.3\%$ as the input MFCC frames were increased from $17$ to $35$ frames; on Wall Street Journal speech corpus. However, this has a major shortcoming -- this information about the adjoining speech frames is being provided as raw MFCC feature frames, which is not handled in a manner that is distinctive from the centred raw speech frames.  On the other hand, the presence of memory cell and recurrent connections in LSTM allows to learn an adaptive and succinct representation, to encapsulate information present in the adjoining speech frames and likely to be relevant for phoneme/triphone classification \cite{graves_hybrid}. 

Our solution to these challenges is moderately inspired by the success of \texttt{word2vec} \cite{mikolov2013distributed} approach, developed in the NLP research community. The \texttt{word2vec} algorithm learns a distributed and compositional representation of text words by optimizing a likelihood function that maximizes the probability of co-occurring words and their respective context words in a text corpus. With similar motivations, we first learn DTE for triphones utilizing frames on a speech corpus, and then use DTE as additional features, along with MFCC features to finally train a second-stage DNN, to output the posterior probability for current speech frame. DTE embeddings are learnt in a discriminative manner and can also be interpreted as an alternate to memory (fixed number of embedding frames are held and utilized latter on) and thereby improve the second-stage DNN's classification accuracy. Similar to the workings of language models, which assign high probability to those words which are likely to co-occur with the context words, the DTEs when used as an additional input features along with MFCC, increase the DNN output probability of  triphone class that is most likely to be true.        

The main contribution of our paper is to encapsulate relevant information from the previous and next speech frames in form of DTE. We train a first-stage four hidden layer DNN with the tied-triphone classification criterion \cite{dahl2012context,DNNKWSTyagi} and utilize the representations learned by its last hidden layer after dimensionality reduction to generate DTE. We then input adjoining DTE frames along with MFCC features to second-stage DNN, to perform the task of tri-phoneme classification. Training both the DNNs on the same objective function (tied-state triphone classification) renders the generated DTE representation aware of the followed task and thus improves the learning of final second-stage DNN. Furthermore, we experiments for phoneme recognition to ensure that the accuracy improved in triphone/phoneme classification helps further for  phoneme recognition.

The remainder of this paper is  as follows: In Section \ref{sec:triphone_embedding}, we describe the proposed method to generate DTE representations. An introduction to hybrid HMM-DNN system for phoneme recognition is presented in Section \ref{sec:baseline}. Experiments and results are provided next in Section \ref{sec:experiments_results}. We summarize our conclusions and suggest future directions of research in Section \ref{sec:conclusion}.

\section{the proposed method}
\label{sec:triphone_embedding}

\begin{figure}[h]
\centering
\includegraphics[width=0.45\textwidth,height=10cm]{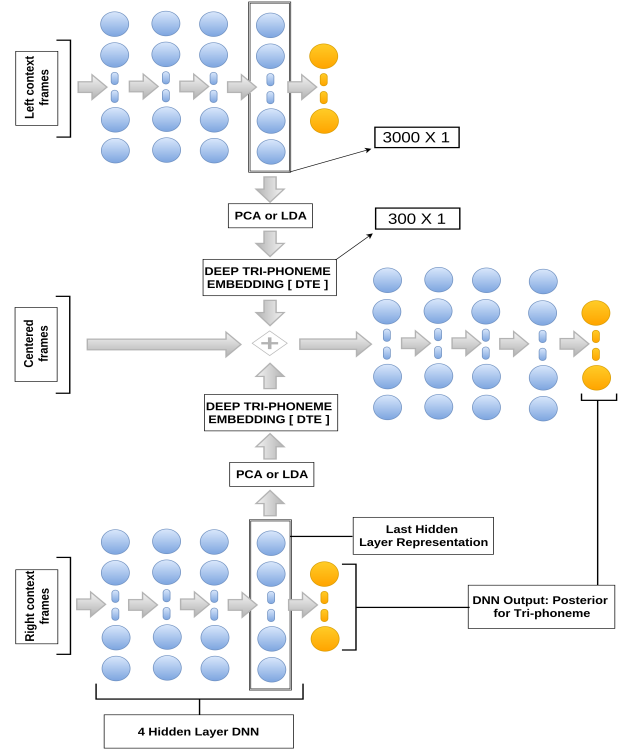}
\caption{The proposed DTE representation based approach.}
\label{fig:dte_schematic}
\end{figure}

The proposed method involves two stages of learning:  i) training and generating a DTE representation for the speech frames arising from both left and right context around few centered speech frames, and ii) the use of  adjoining frames' DTEs along with MFCC feature vectors of few center frames as input to second-stage DNN. This approach is illustrated schematically in Fig.\ref{fig:dte_schematic}. More formally, one can describe this task as to learn a mapping function $\mathtt{f}$, between tied-triphone labels and the MFCC feature vectors arising from left context, center frames, and right-context frames ($[x_l,x,x_r]$) of speech, as mentioned below in Equation \ref{eq:dte}:

\begin{equation}
\label{eq:dte}
\mathtt{f} ([x_l,x,x_r]) = \mathtt{ g } ( [ \mathtt{ h}  ( \mathtt{ \overline{ g } }(x_{l})) , x , \mathtt{ h } ( \mathtt{ \overline{ g } } (x_{r})) ] )  
\end{equation}

where $\mathtt{ \overline{g} }$, $\mathtt{ g }$, represent first-stage DNN and learnt function of second-stage DNN's respectively.  Here, $\mathtt{h}$ denotes a functional operator which extracts the last hidden layer's activation vector ($3000$ dimensional corresponding to each node in last hidden layer) of the first-stage DNN. After that we reduces  dimension through Principal Component Analysis (PCA), to obtain the final $300$ dimensional vector; which is the succinct representation, named as DTE. 

We next describe the reasons why we choose the DNN's last hidden layer's activation vector as the DTE. Consider a DNN with $G$ hidden layers (where $G=4$ in this paper) and with output vector $y(t)$, which models the class\footnote{Tied-state triphones are used as output classes.} posterior probability given the input $x(t)$ (MFCC features). This is achieved by using a \textit{softmax} non-linearity at the output layer. Whereas the nonlinearity at hidden layers nodes could be \textit{tanh}, or Rectified Linear Units (\textit{ReLU}). Posterior probability that input feature $x(t)$ belongs to class $'m'$, is:
\begin{equation}\label{e3b}
\begin{split}
y_m(t) & = P( x(t)\: \in \:m) | x(t)) \\
& = \frac{\exp( \sum_{j=1}^{N_G} w_{m,j}^{G+1}z^G_j(t) ) }{ \sum_{k=1}^M\exp( \sum_{j=1}^{N_G} w_{k,j}^{G+1}z_j^{G}(t) )   } 
\end{split}
\end{equation}
where $ w_{m,j}^{G+1}\: j\in(1,N_{G})$ is weight vector connecting $m^{th}$ output unit to all the hidden units in the last hidden layer $G$, which has $N_{G}$ hidden units. As explained in \cite{dl_book,hinton2012deep}, the workings of a DNN can also be interpreted as follows:

\begin{itemize}
\item A DNN with $G$ hidden layers maps the input feature $x(t)$ to a high-dimensional and sparse non-linear feature space $z(t)$ which is the activation vector at the $G^{th}$ hidden layer. This new feature space $z(t)$ makes the classification problem easier to solve than the original raw feature space $x(t)$.
\item Then, the \textit{Softmax} non-linearity at the output layer can also be interpreted as a set of Logistic Regression based classifiers; acting on the last hidden layer's activation vector ($z(t)$) and exploiting the mapping learnt in the hidden layers to generate new and effective features.

\end{itemize}

Based on this reasoning, we retain activation vector at the last hidden layer of the first-stage DNN and followed by dimensionality reduction as the \textit{Deep Triphone Embedding (DTE)} - as a representation for previous and next frames to provide relevant information for the task of phoneme classification. The activation vectors at the hidden layers are known to be highly sparse and hence we could further compress them to a low -dimensional representation by using either a Principal Component Analysis (PCA) or Linear Discriminant Analysis (LDA) transform. Dimensionality reduction also helps in reducing the number of model parameters for the second-stage DNN. Both the first-stage DNN $\mathtt{ \overline{ g } }$ and second-stage DNN $\mathtt{ g }$ are trained using the cross-entropy loss; which is defined with respect to tri-phoneme labels obtained after force alignment using a well trained tied-state triphone HMM-GMM acoustic model.

\section{HYBRID HMM-DNN SYSTEM}
\label{sec:baseline}

The speech recognition performance have improved tremendously when the DNN is trained on the tied-state triphones states as the output labels than the monophone labels \cite{dahl2012context,DNNKWSTyagi}. This is due to the improved co-articulation modelling by the triphone states. However, the triphone states are much larger in number than the monophone states ($1373$ and $40$ respectively in our experiments) and training a DNN with triphones as labels is a harder learning problem than training with the monophone labels.    

We have trained a strong baseline tied-state triphone GMM-HMM acoustic model with $1373$ tied states on a subset of TED-LIUM corpus \cite{ted_lium}. TED-LIUM corpus consist of $1495$ TED talks given in English language by different speakers covering a wide variety of speaking styles, and demographic and linguistic backgrounds (US English, British English, Continental European English, Indian English, Chinese English and Australian English speakers), making it a challenging multi-accented conversational speech corpus for speech recognition. We have developed a C++ library to train CD-HMM-GMM, which uses the standard Expectation-Maximization (EM) algorithms for the parameter estimation and a decision tree based tying that resulted into $1373$ tied-triphone states in our experiments. This system is also used to produce tied-state triphone labels for the entire train, dev and test-sets via forced alignment in order to train our $4$ hidden layer DNNs. We use the traditional $24$ Mel filter bands, followed by $13$ DCT coefficients of the log Mel filterbank energies and their delta and delta-delta coefficients, resulting into $39$ dimensional MFCC feature vector for each speech frame. This DNN is provided a context of $2\times P+1$ adjoining MFCC frames (center frame $\pm P$ frames), resulting into $39\times (2\times P+1)$ dimensional input feature. The network processes the input through a sequence of ReLU non-linearities \cite{glorot2011deep}. In particular at the $i^{th}$ layer, the network computes: $\textbf{h}^i=f(\textbf{W}^i\textbf{h}^{i-1}+\textbf{b}^i)$, where $ \textbf{W} \in \mathcal{R}^{M \times N}$ is a matrix of trainable weights, $\textbf{b}^i \in \mathcal{R}^M$ is vector of trainable biases, and   $\textbf{h}^i \in \mathcal{R}^M$ is the $i^{th}$ hidden layer and  $\textbf{h}^{i-1} \in \mathcal{R}^N$ is the $(i-1)^{th}$ hidden layer (or the input $x$ if $(i-1)=0$). 

\begin{figure}[h]
	\centering
	\includegraphics[width=0.5\textwidth,height=8cm]{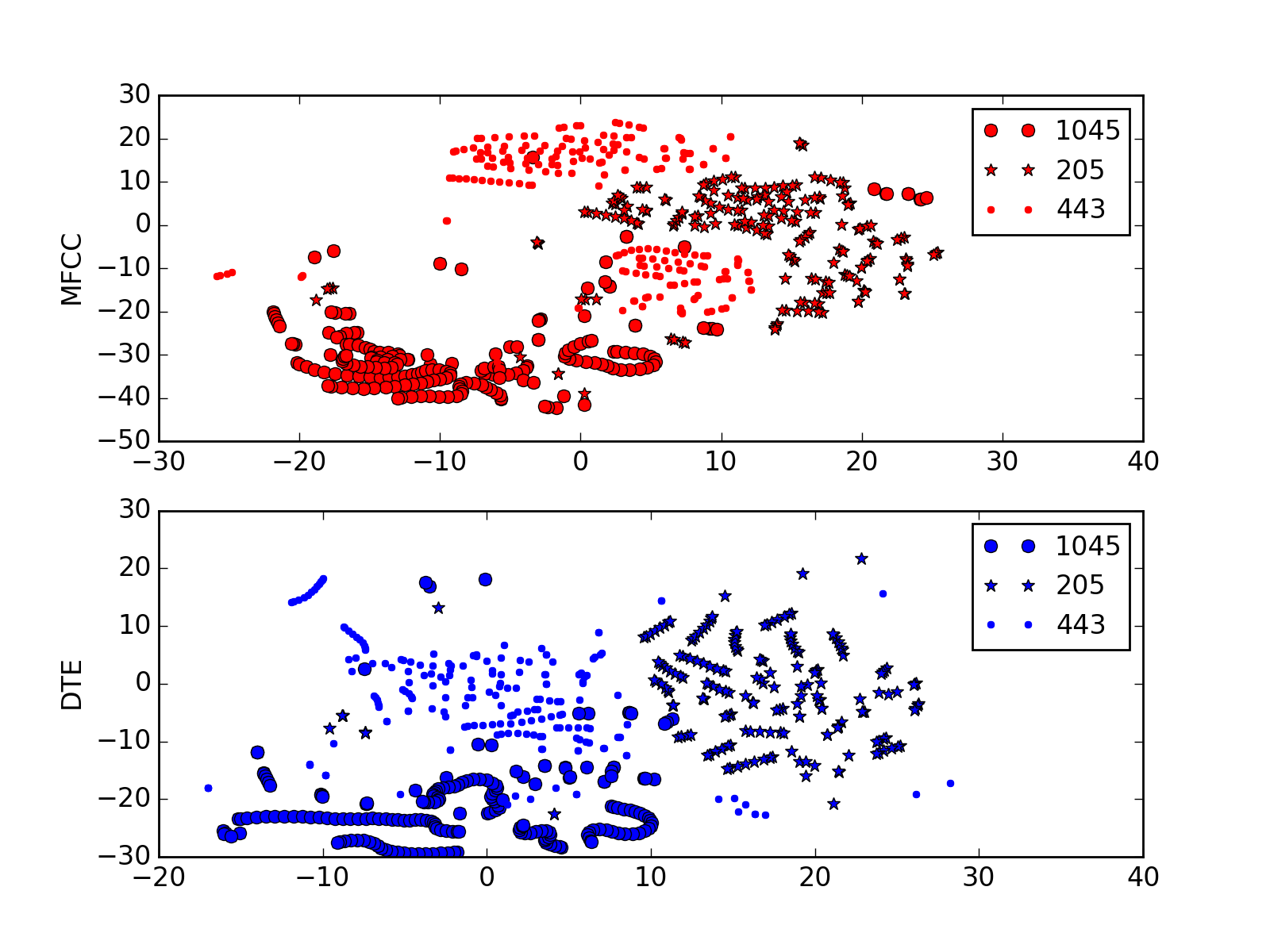}
	\caption{Visualizing of three tied-states (decision tree labels : 1045, 205, 443) using t-SNE based dimension  reduction on: a) raw MFCC features for 45 centred frames (upper panel) and b) centred MFCC features for 21 frames + DTE of 15 previous (lower panel).}
	\label{fig:dte_tsne}
\end{figure}

Theano library \cite{bastien2012theano} is used to train the DNNs using Stochastic Gradient Descent (SGD) with progressively decaying learning rate which starts at $0.01$ and is validated with a small dev-set to reduce the learning rate. All parameters in the weight and bias matrices are initialized randomly. 

\section{EXPERIMENTS AND RESULTS}
\label{sec:experiments_results}
\subsection{Dataset Description}

\noindent Experiments were performed on the $300$ talks subset of the TED-LIUM corpus which originally consists of $1495$ TED talks\cite{ted_lium}. We took the first $300$ talks from this data-set to form our train-set and dev-set of $40$ and $2$ hours duration respectively (these durations are after removing excessive silence frames which dominate in any speech corpus). We use the official TED-LIUM test-set ($2Hrs$ of speech and consists of $11$ talks).  We use the official pronunciation dictionary which consists of about $152K$ words and comprises of $40$ phonemes/monophones. The baseline tied-state triphone HMM-GMM system has $1373$ tied-states, and each triphone is modeled by three state left-to-right HMM with a GMM based emission distribution with $11$ Gaussian mixture components. We used our customized C++ library for estimating these parameters through EM algorithm and decision tree based state tying. 

\subsection{Phoneme Classification \& Recognition}

\noindent We have used the optimal Viterbi decoding with all the acoustic models (HMM-GMM,  HMM-DNN variants, \& proposed methods), where in case of the HMM-DNN acoustic model, scaled likelihoods, which are just the posterior probabilities of tied-triphone states obtained at the DNN output layer, are used for the triphone state likelihoods.  A bigram phoneme language model is learned on the train-set phoneme transcripts. The LM factor and phone insertion penalty are empirically tuned on the dev-set. It is worth emphasizing that we measure performance only on non-silence frames for both the phoneme classification and recognition experiments; as silence frames are dominating (typically $10\%$) in a speech corpus  and can inflate the results.  All of our baselines and the proposed DTE based methods are trained for the tied-triphone labels. In order to obtain the phoneme classification and recognition accuracies, we map the center phone of triphone label as the predicted phoneme.

\begin{table}
	\centering
	\small{
		\begin{tabular}{|c|c|c|c|c|}
			\hline
			{Training Methods} & \multicolumn{1}{c|}{Tri-Phoneme} & \multicolumn{1}{c|}{Phoneme} \\ 
			\hline
			HMM + GMM          	    &   17.63         & 41.21                 \\ \hline
			HMM + DNN             	&   29.30         & 62.52                 \\ \hline
			HMM + DNN - W            	&   36.09         & 66.05                 \\ \hline
			HMM + DNN - W+D          	&   33.75         & 64.31                 \\ \hline
			HMM + DTE-LDA + DNN 	&   \bf{41.40}   &       \bf{68.27}           \\ \hline
			HMM + DTE-PCA + DNN 	&	\bf{42.27}           &   \bf{68.31}              \\ \hline

		\end{tabular}
		
	}
	\caption{ Tri-phoneme and phoneme classification accuracy.}
	\label{table:classification}
\end{table}

\begin{table}[]
	\centering
	\begin{tabular}{@{}|c|c|@{}}
		\hline
		Training Methods & Phoneme accuracy \\
		\hline
		HMM + GMM                  &  55.72                \\ \hline
		HMM + DNN                &          63.50        \\ \hline
		HMM + DNN - W               &           68.11       \\ \hline
		HMM + DNN - W+D               &          66.11        \\ \hline
		HMM + DTE-LDA + DNN              & \bf{ 70.11}               \\ \hline
		HMM + DTE-PCA + DNN             &   \bf{ 70.22 }             \\
		\hline
	\end{tabular}
	\caption{Phoneme recognition accuracy.}
	\label{table:recognition}
\end{table}

For comparison, we have prepared four baselines, namely, HMM+GMM, HMM+DNN, HMM+DNN-W and HMM+DNN-W+D. First one is based on traditional CD-HMM-GMM based generative systems and remaining three are variants of HMM+DNN type od system. HMM+DNN, HMM+DNN-W and HMM+DNN-W+D are with P=10 \& 4 hidden layers, P=24 \& 4 hidden layers, and P=24 \& 8 hidden layers; where P represents number of previous or next frames used. Last two baselines are as wide as our method and this is to ensure that the improvement seen in proposed is not merely due to the high number of previous or next frames consumed. The last baseline which is HMM+DNN-W+D, is  made of 8 hidden layers and ensures that the improvement is not achieved merely due to increase in the partial depth of DTE in the proposed methods.

 Table \ref{table:classification} shows the results for tri-phoneme and phoneme classification with various methods used to train an acoustic model. The performance improvement by HMM+DNN variants based acoustic model over previous HMM+GMM method validates the discriminative strength of DNN based models. On the top of that, both the variants of our proposed DTE based methods, namely, HMM+DTE-LDA+DNN and HMM+DTE-PCA+DNN achieves an absolute improvement of $\mathtt{5.31\%}$ / $ \mathtt{6.18\%} $ and $\mathtt{2.22\%} $ / $ \mathtt{2.26\%}  $, in tri-phoneme / phoneme classification accuracy, respectively. The relative improvements of $\mathtt{8.31\%}$ and $\mathtt{6.66\%}$ in tri-phoneme / phoneme classification accuracy are achieved by HMM+DTE-PCA+DNN, when compared against all variants of HMM+DNN based baselines.

The results on phoneme recognition are presented in Table \ref{table:recognition}, for various training methods used to learn the acoustic models. We report phoneme recognition accuracy by taking into account all the \textit{substitution, deletion and insertion} errors. The performance of variants of the proposed methods, namely, HMM+DTE-LDA+DNN and HMM+DTE-PCA+DNN indicates that the proposed method delivers superior performance, hence offers a huge potential to build efficient and robust speech recognition systems. More precisely, variants of proposed methods achieve as high as an absolute improvement of $ \mathtt{2.11\%}$ and relatively improvement of $ \mathtt{6.61\%}$; when compared with a very strong baseline. It is important to note that the further optimization of dimension of DTE might results in further improvement which was fixed to 300 in our experiments. 

To investigate the effect of the proposed DTE representation closely, we compute DTE representation corresponding three tied-state triphone labels [1045,205,443] generated by decision tree. The DTE representation was reduced to two dimension using t-SNE method to be able to visualize them \cite{t_sne}; as shown in Figure \ref{fig:dte_tsne}. Visualization makes it clearly apparent that on addition of DTE representation, two of the more confounded triphone ([205,443]) are much more disentangled, in comparison to only raw MFCC features.

\section{RELATED WORK}
\label{sec:related_work}

Deep neural networks have played an important in the resurgence of newly advanced speech recognition systems \cite{deep_speech_2}. One of the popular direction of research is to train neural networks in an end-to-end fashion, aimed to generate transcriptions using RNN \cite{ctc}. These methods enjoy the advantage of direct training over target linguistic sequence conditioned on the input acoustic sequence and do not require an explicit step of force alignment to segment the acoustic data. However, these methods stumble to integrate with large vocabulary speech recognition systems; due to their inability to combine easily with word-level models \cite{graves_hybrid}. Contrary to these methods, hybrid HMM-GMM based methods can embed word-level information seemingly; which is vital in developing real-world systems. Hence, the focus of this paper limits to only hybrid HMM-GMM systems. 

Difficulty in training deep models has been one of the most critical factor to impede in widely spreading their usage until very recently. Few of the most critical challenges include but not limited to, initializing weights of the network, choice of optimal algorithm for optimization and hyper-parameters, and regularizing to achieve higher generalization performance \cite{icml_hinton}. Specially, training RNNs is a hard problem; when compared with DNNs for multiple reasons \cite{advances_rnn_optimization,recurrent_batch_normalization,rnndrop}. Inspired from these issues, this paper seeks for a method based on only DNNs. Though, in past, RNNs has been studied quite extensively for speech recognition \cite{phoneme_classification_2,graves_hybrid,ctc,deep_speech_2,rnndrop}. Research studies on RNNs has revealed that bidirectional LSTM recurrent neural networks can be utilized as an acoustic model in a manner similar to hybrid HMM-DNN systems. However, the improvement in recognition accuracy over DNNs is reported to be modest \cite{graves_hybrid}. Interestingly, bidirectional nature of these networks can capture both left and right contextual frames of speech, similar to this paper. 

Recently, highway networks  have been also proposed; motivated by the fact that the network depth is crucial to learn better models \cite{highway}. While highway networks looks conceptually similar, this work does not  advocate information flow across through direct connections between the layers of network. Although, our method also enjoy the advantage of selective higher depth; as in data from previous and next frames go through 8 hidden layers effectively.

\section{CONCLUSION AND FUTURE WORK}
\label{sec:conclusion}

RNN and more specifically LSTM\cite{graves_hybrid}, owing to their explicit memory cell, have shown great promise in sequence recognition tasks. Traditional DNN  does not have such a memory cell to persist useful leaned information from the adjoining speech frames, and hence is at a disadvantage \cite{dahl2012context}. However, training a DNN is relatively easier as compared to training and tuning a LSTM which has complex interactions of the various input, output, and memory gates which pose significant challenges in SGD based parameter learning \cite{difficult_rnn}.     

This paper has presented a novel DTE representation which augments the classical DNN and imparts it partial LSTM like capability by enabling them to retain information from context; without comprising on the simplicity of the DNN training procedures. The proposed method yields superior improvements of absolute $\mathtt{6.18\%}$, $\mathtt{2.26\%}$, and $ \mathtt{2.11\%}$ in tri-phoneme classification, phoneme classification, and phoneme recognition respectively; when compared the baseline classical hybrid HMM-DNN system. 

Finally, we note that the results offered in this work are achieved on the challenging TED-LUM \cite{ted_lium} corpus which consists of multi-accented English TED talks given by speakers from a very diverse demographic and linguistic backgrounds (US English, British English, Continental European English, Indian English, Chinese English and Australian English speakers). In future, we would investigate and extend our analysis with the full set of available data ($207$ hours) and Large Vocabulary Speech Recognition experiments. 

\bibliographystyle{IEEEbib}
\bibliography{Template}

\end{document}